\documentclass{article}

\usepackage{arxiv}

\usepackage[utf8]{inputenc} 
\usepackage[T1]{fontenc}    
\usepackage{hyperref}       
\usepackage{url}            
\usepackage{booktabs}       
\usepackage{amsfonts}       
\usepackage{nicefrac}       
\usepackage{microtype}      
\usepackage{lipsum}		
\usepackage{graphicx}
\usepackage{natbib}
\usepackage{doi}
\DeclareUnicodeCharacter{2264}{$\leq$}

\title{Observation of Alfv\'en waves in an ICME-HSS interaction region }
\author{Omkar Dhamane$^1$, Anil raghav$^{1*}$, Zubair Shaikh$^2$, Utsav Panchal$^1$, Kalpesh Ghag$^1$,\\ Prathmesh Tari$^1$, Komal Choraghe$^1$, Ankush Bhaskar$^3$, Daniele telloni$^4$, Wageesh Mishra$^5$ \\
\\
$^1$Department of Physics, University of Mumbai, Mumbai, India\\
$^2$Indian Institute of geomagnetism, Panvel, Navi Mumbai, India\\
$^3$ Vikram Sarabhai Space Centre (VSSC), Indian Space Research Organisation (ISRO), Thiruvananthapuram, Kerala 695022, India\\
$^4$National Institute for Astrophysics, Astrophysical Observatory of Torino, Via Osservatorio 20, I-10025 Pino Torinese, Italy\\
$^5$Indian Institute of Astrophysics, II Block, Koramangala, Bengaluru 560034, India \\
	\texttt{*anil.raghav@physics.mu.ac.in} \\ }
   
\renewcommand{\shorttitle}{Dhamane et. al. }

\hypersetup{
pdftitle={A template for the arxiv style},
pdfsubject={q-bio.NC, q-bio.QM},
pdfauthor={David S.~Hippocampus, Elias D.~Striatum},
pdfkeywords={First keyword, Second keyword, More},
}

\begin{document}
\maketitle
\begin{abstract}
 The Alfv\'en wave (AW) is the most common fluctuation present within the solar wind emitted from the Sun. Whether or not AWs can originate after the collision of an Interplanetary Coronal Mass Ejection (ICME) and a High-Speed Stream (HSS) remains an open question. To find an answer to this question, we have investigated an ICME-HSS interaction event observed on $21^{st}$ October 1999 at 1 AU by the WIND spacecraft. We have used the Wal\'en test to identify AWs and estimated the Els\"asser variables to find its characteristics. We explicitly find dominant sunward AWs within the ICME whereas the trailing HSS had strong anti-sunward AWs. We suggest that the ICME-HSS interaction deforms the Magnetic Cloud (MC) of the ICME, resulting in the generation of AWs inside the MC. Additionally, the existence of reconnection within the ICME's early stage could also contribute to the origin of AWs within it. 

\end{abstract}

\section{Introduction}
     \label{S-Introduction} 

A coronal mass ejection (CME) is a massive expulsion of a considerable volumes of plasma with immense energy flows  from the solar corona \citep{hundhausen1999coronal, webb2012coronal}. CMEs, High-speed stream (HSS), and Corotating interaction regions (CIRs)  are the primary sources of severe space weather conditions in the heliosphere and planetary environments \citep{tsurutani2006corotating,tsurutani2006recurrent,schrijver2010heliophysics}. A CME's relative excess speed over the ambient solar-wind speed forms the shock front and sheath region \citep{tsurutani1988origin, tsurutani2011review, kilpua2017coronal}. 
 Fast-forward shocks generate upstream of the CME, i.e., ahead of the solar-origin plasma and field structures \citep{kennel1985quarter,tsurutani2011review}. 
When a CME moves from the near-Sun region to the interplanetary medium, its kinematic configuration possibly evolves \citep{vrvsnak1993kinematics}. 
ICMEs are the Interplanetary counterparts of CMEs observed in the heliosphere using in-situ data. ICMEs cause extreme geomagnetic storms and disruption in the heliosphere and magnetosphere \citep{tsurutani1992great,  zurbuchen2006situ, echer2008interplanetary, zhao2014current, kilpua2017geoeffective, meng2019solar}. Their hazardous impacts on Earth's space weather are complex for current spacecraft technology to handle \citep{board2009severe}. ICME research has received a lot of interest due to its importance in scientific and technical implications \citep{schrijver2010heliophysics, webb2012coronal,cannon2013extreme}.  

The Solar wind was first referred to as the phenomenological ``solar corpuscular radiation" that causes
	geomagnetic and auroral activity \citep{parker1965dynamical}. Generally, it is distinguished into two classes: fast solar wind (speed $>400$ km s$^{-1}$) and slow solar wind (speed $<400$ km s$^{-1}$)\citep{belcher1971large, dasso2005anisotropy, feldman2005sources, abbo2016slow, tsurutani2022extremely}. The wind's speed is not the only parameter to differentiate between the two, but their relative composition also characterizes the steady bulk plasma properties \citep{von2000composition}. The fast solar wind originates from  coronal holes (CHs) in the Sun 
	\citep{krieger1973coronal, gosling1999formation, vrvsnak2007coronal}. In addition, data from Ulysses indicates that coronal hole high-speed streams move at a speed of 750–800 km s$^{-1}$ \citep{balogh1995heliospheric, mann2000interstellar}.
 When an ICME travels in the heliosphere through the solar wind, their interaction, particularly with HSSs, can significantly affect the ICME's properties. This results in the ICME's embedded flux rope to bend, kink, rotate, or become distorted \citep{riley2004kinematic, manchester2004modeling, wang2006impact}. 
 Sometimes, the flux rope gets eroded due to reconnection in the ICME-HSS interaction \citep{dasso2006new, ruffenach2012multispacecraft, lavraud2014geo, ruffenach2015statistical}.


The solar wind, particularly the HSS, is characterized by large amplitude Alfv\'en waves \citep{belcher1971large, tsurutani1995interplanetary, tsurutani1996interplanetary, tsurutani2017alfvenic, tsurutani2018review}. It is a primary fluctuation in a magnetized plasma, notably incompressible magnetohydrodynamic (MHD) waves \citep{alfven1942existence}. Due to their distinct properties, Alfv\'enic oscillations have been regarded as the best MHD mode for energy transport \citep{alfven1947granulation, mathioudakis2013alfven}. The Sun is considered as the primary source of AWs \citep{belcher1971large}. It is hypothesized that the magnetic reconnection or catastrophe processes that occur during the onset of a CME, may result in the generation of low-frequency anti-sunward  AWs and fast- and slow-mode magnetoacoustic waves  \citep{kopp1976magnetic,antiochos1999model, chen2000emerging}. However, sunward propagating AWs indicate
	their generation in interplanetary space.
	The origin of such AWs is associated with several physical processes such as: magnetic reconnection exhausts \citep{belcher1971large,gosling2009one}, backstreaming ions from reverse shocks \citep{gosling2011pulsed}, steepening of a magnetosonic wave \citep{tsurutani2011review,tsurutani1988origin}, interaction of multiple CMEs \citep{raghav2018first}, etc. It has been proposed that AWs are produced locally by velocity shear instabilities caused by their interaction with high-velocity streams \citep{coleman1968turbulence, bavassano1978local, roberts1987origin, roberts1992velocity} . Besides, AWs are locally generated due to kinetic instabilities linked to the solar-wind proton heat flux \citep{neugebauer1981observations}. According to models of the expanding solar wind by \cite{matteini2006parallel}, and \cite{hellinger2008oblique}, the plasma is unstable to the firehose and oblique firehose instabilities at a distance of about 1 AU. Intriguingly, oblique AWs are produced by the firehose instability. Additionally, the solar wind expansion increases the ratio of differential particle velocity to local Alfv\'en speed, which leads to the oblique AWs instability \citep{hellinger2011proton, hellinger2013protons}. 

AWs play a crucial role in both the dynamics of the Earth's magnetosphere and the study of space plasmas in astrophysics. (e.g., \citep{cummings1969standing,singer1981alfven, tsurutani1987cause, hui1992electron,johnson1997kinetic, lysak2004magnetosphere} and references therein). \cite{tsurutani1987cause} claimed that high-intensity long-duration continuous auroral activity (HILDCAA) events are caused by outward (from the Sun) propagating interplanetary AW trains. The HSSs contain AWs linked to HILDCAA events \citep{tsurutani2011properties}. 
	 Moreover, \cite{chaston2007important}  hypothesized that auroral particles can be accelerated by AWs. The Alfv\'enic fluctuations or AWs can affect the ionosphere, resulting in various phenomena \citep{verkhoglyadova2013variability}. The ponderomotive force generated by an AW can create a cavity into the ionospheric plasma \citep{bellan1998fine}. \cite{chaston2006ionospheric} hypothesized that AWs cause ionosphere erosion. \cite{hull2019dispersive} suggested that the dispersive AWs become quite essential in energising $O^+$ ions in the inner magnetosphere. Moreover, during geomagnetic storms, the Alfv\'enic fluctuations inside the ICME substructures can cause the extended recovery of the geomagnetic field  \citep{raghav2018torsional,raghav2019cause,shaikh2019concurrent,telloni2021alfvenicity}. Besides, it was also shown that during the storm's main phase, the hemisphere's  Alfv\'enic power surged four times when compared to non-storm periods \citep{keiling2019assessing}. In addition,  AWs play a significant role in plasma heating \citep{hasegawa1974plasma}, transportation (e.g., \citep{hasegawa1975kinetic,chen2016physics} and references therein), magnetotail dynamics \citep{keiling2009alfven}, auroral dynamics \citep{stasiewicz2000small}, etc. Therefore, it is important to investigate the origin and propagation of the AW and its related processes.

The structural configuration of  large-scale magnetic structures is altered by interactions, such as CME-CME, CME-HSS, CME-CIR etc. \cite{ heinemann2019cme} 
showed the signature of a Stream interface(SI) as the HSS passes the slow solar wind, resulting in a drop in proton density and a sharp increase in temperature. They further pointed out that the HSS follows the CME, and their interaction gives a sharp rise in the magnetic field, proton density, velocity and temperature corresponding to the shock sheath region. Theoretical studies also suggest that when large-scale magnetic structures interact, momentum and energy are transferred in the form of an MHD wave \citep{jacques1977momentum}. \cite{raghav2018torsional} investigated a CME-CME interaction event and discovered torsional AWs in the magnetic cloud (MC) region. Moreover, Alfv\'enic fluctuations were found in the ICME sheath \citep{raghav2022situ} and the stream interface of fast and slow solar-wind interaction \citep{lepping1997wind,tsurutani1995large}. Here, we present the first observation of AWs generation caused by the interaction between the ICME and following HSS.

\section{Data, Methods and Observations }

We examined the ICME-HSS interaction event observed by the WIND spacecraft on $21^{st}$  October 1999. We used 92 sec temporal resolution data of the plasma and magnetic field in GSE coordinates to examine the interplanetary conditions during the passage of the interaction region. Moreover, to determine the presence of AWs, we analysed high-resolution data from WIND's satellite sensors (such as WIND MFI, and 3DP) with a 3 sec resolution. The data is available at \url{wind.nasa.gov/data.php}.

\subsection{Interplanetary Conditions}

\begin{figure}
    \centering
   \includegraphics[width = \columnwidth]{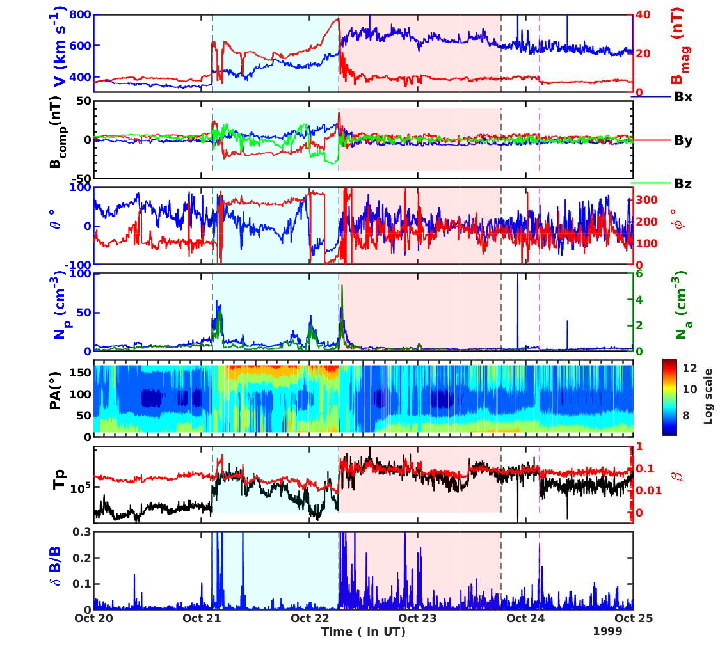}
   \caption{Wind observation of complex ICME–HSS interaction event on 1999 October 20–24 (time cadence of 92 sec) Total interplanetary field strength IMF $B_{mag}$ in nT and total solar wind V (km s$^{-1}$) are shown in the top panel. The components of the magnetic field are shown in the second panel. The third panel displays the IMF orientation ($\theta$, $\phi$). In the fourth panel, the proton number density ($N_p$) is represented on the left, while alpha density $N_a$  is shown on the right. The pitch angle (PA) of superthermal electron strahls is depicted in the fifth panel. The proton temperature ($T_{\rm P}$) and the $\beta$ value were plotted in the sixth panel on the left- and right- sides, respectively. The plot of  $\delta B$/B is demonstrated in last panel}
   \label{fig:IP_label}
\end{figure}

The interplanetary conditions during the passage of the ICME-HSS interaction region are demonstrated in Figure ~\ref{fig:IP_label}. A sudden sharp enhancement is observed in the total interplanetary magnetic field (IMF), plasma density, and solar-wind speed, suggesting the onset of the ICME at 02:19 UT on 21 October 1999. The low plasma beta ($\beta$) and low fluctuations in the IMF indicate the magnetic cloud (MC) crossover. The electron pitch-angle shows a nearly bidirectional flow that confirms a possible closed magnetic structure. The rear-end is observed at 06:29 UT, on 22 October 1999. There are also signatures of reverse shock on the 24 October 1999 at 03:04 UT, i.e., decrease in magnetic intensity, temperature, and number density (indicated by a dotted magenta line in Figure ~\ref{fig:IP_label}). The ICME boundaries are also confirmed by the ICME catalogue available at \url{wind.nasa.gov/ICME_catalog/ICME_catalog_viewer.php }

In general, MCs of ICMEs depict a gradual decrease in total IMF and solar-wind speed, which implies the expansion of the MC in the solar wind. However, in the studied event, the trailing edge of the MC demonstrates an anomalous behaviour, such as a rise in the total IMF and solar-wind speed. Thus, it is clearly visible that the observed ICME's MC completely contradicts the conventional definition of a MC. The anomalous behaviour at the trailing edge of the ICME's MC could be due to the existence of a HSS flow from behind. The compression of the ICME's MC by the HSS causes the plasma particles to pile up at the trailing end, along with the high IMF fluctuations  \citep{rodriguez2016typical}. The temperature and density rise at the interface between an ICME and HSS causes thermal pressure to rise as well. As a result, 
the force balance conditions of the ICME flux rope may be altered. Hence, we believe that the MC is distorted due to the ICME-HSS interaction, and the corresponding HSS may cause turbulence at the rear end of the MC. Additionally, at the leading part of the ICME, we observe two sharp dips in the total magnetic field, which coincides with a rise in proton density, alpha density and plasma temperature. This can be interpreted as magnetic reconnection exhaust, based on reported literature \citep{gosling2005magnetic}.

\begin{figure}
	\centering
	\includegraphics[width = \columnwidth]{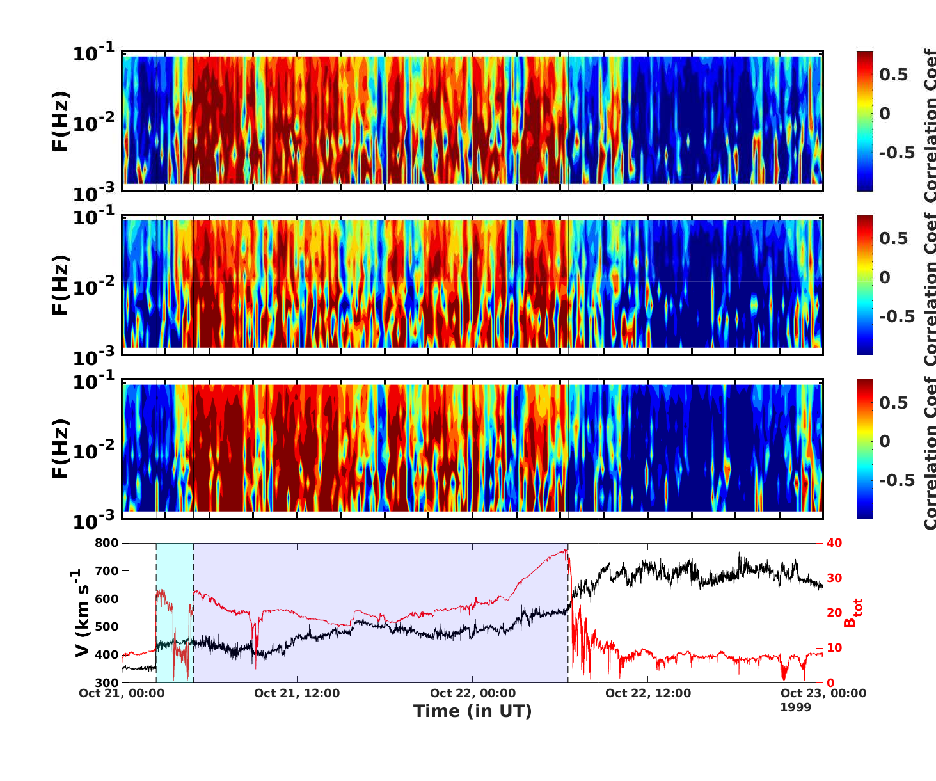}
	\caption{In the time frequency domain, the correlation coefficients between $V_{Ai}$ and $V_i$ for the ICME-HSS event are presented. The top three panels show the correlation coefficient for $x$, $y$, and $z$ components, respectively.  The bottom panel shows the total magnetic field ($B_{\rm mag}$), and solar-wind velocity (V) is plotted for reference. The shaded region in the bottom panel depicts ICME region.}
	\label{fig:wal_label}
\end{figure}
 \subsection{Alfv\'en Wave Identification}
 
AWs are the most basic form of fluctuation in a magnetic plasma, commonly identified in the solar wind across the heliosphere \citep{alfven1942existence,belcher1971large}. 
 The Alfv\'en velocity fluctuations $\Delta V_A$ are defined as,
\begin{equation}
	\Delta V_A=\frac{\Delta B}{\sqrt{\mu_0 \rho}}
\end{equation}
where, $\Delta B = B - B_{avg}$ are the fluctuations in the respective components of the IMF. A correct assessments of the fluctuations magnitude requires the accurate estimate of background values. In the literature, a precise de Hoffmann-Teller frame or mean values are utilized as a background value to diagnose the existence of interplanetary AWs
\citep{gosling2009one, yang2013alfven,raghav2018first,raghav2018torsional}.  However, in an HSS, the HT frame might vary rapidly \citep{gosling2009one,li2016new}, and the use of an average value as the background state is not always appropriate. 
As a result, during the examined ICME-HSS interaction, we employed other techniques to identify large-amplitude AWs. The entire data set being analyzed is divided into ten-minute time intervals. Each time window's data is passed through the $4^{th}$ order Butterworth filter (using MATLAB software). We choose 10 periodic bands for bandpass filter: 10-15 sec, 15-25 sec, 25-40sec, 40-60sec, 60-100 sec, 100-160 sec, 160-250 sec, 250-400 sec, 400s-630 sec, and 630-1000 sec in an evenly distributed manner. In our study, the AWs are observed in frequency bands between $10^{-3}$ to $10^{-1}~$Hz. The Wal\'en relation is used to determine the relationship between the Alfv\'en and the solar-wind velocity components as,
\begin{equation}
	\Delta V_i ~= ~|R_{wi}| ~\Delta V_{Ai}
\end{equation}
where $R_{wi}$ is known as the Wal\'en slope which represent the linear relationship between $\Delta V_{Ai}$ and $\Delta V_i$. Furthermore, the presence of AWs or Alfv\'enic fluctuations in the examined region was demonstrated by the Pearson correlation coefficient between the respective components of $\Delta V_i$ and $ \Delta V_{Ai}$. Figure ~\ref{fig:wal_label} demonstrates the existence of AWs during the ICME-HSS interaction. The colour bar shows the correlation coefficient between the components of  $\Delta V_A$ and $\Delta V$. 
Here, correlation coefficient = -1 (dark-blue shade) implies the existence of anti-sunward AWs, whereas correlation coefficient = 1 (dark-red shade) means sunward AWs. The top three panels show the frequency-dependent distribution of correlation coefficient between the respective components ($x$, $y$, and $z$) of $\Delta V_A$ and $\Delta V$. A negative correlation coefficient was observed at the ICME upstream solar wind and trailing HSS region. In contrast, during the ICME transit, we mainly observed a positive correlation coefficient. This implies the ICME is superimposed with the dominant sunward AWs, whereas the trailing HSS shows anti-sunward flow.


\begin{figure}
	\centering
	\includegraphics[width = \columnwidth]{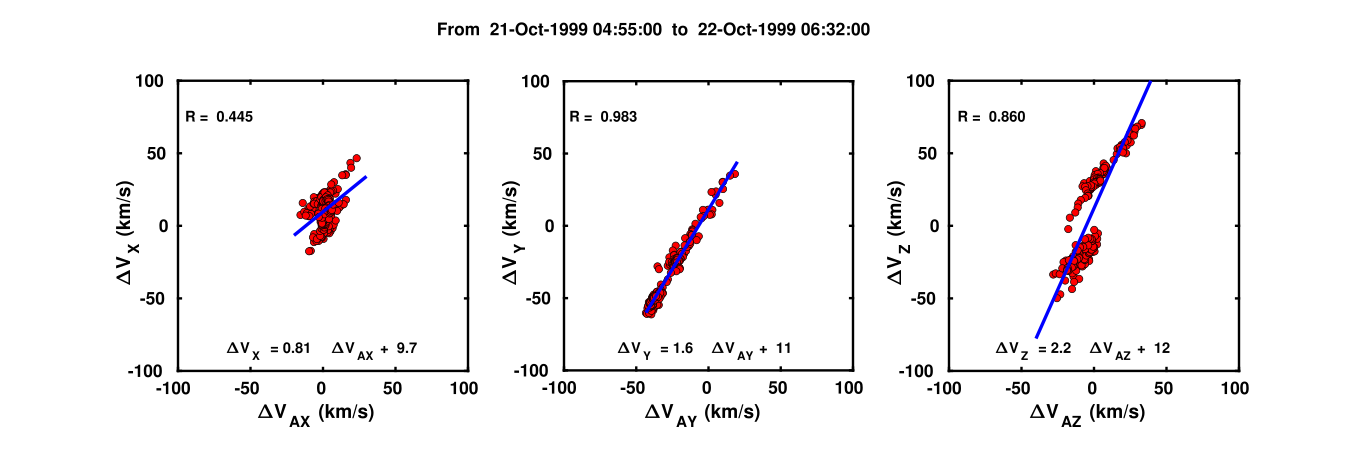}
	\caption{Analysis of the correlation between the corresponding $\Delta V$ and $\Delta V_A$ components. The scattered black circle with filled red colour represents the WIND spacecraft observations with a time cadence of 3 sec. The coefficient of correlation is denoted by R. Each panel shows the linear fit relationship between the corresponding components of $\Delta V$ and $\Delta V_A$.}
	\label{fig:cloud_label}
\end{figure}
Figure ~\ref{fig:cloud_label} depicts the correlation analysis between the respective components of $\Delta V$ and $\Delta V_A$ for ICME's MC. We used the $4^{th}$ order butter-worth MATLAB filter algorithm with a single broadband frequency boundary of $10^{-3}$ to $10^{-1}$ Hz to filter the $\Delta V$ data and $\Delta V_A$ components. We found the Pearson correlation coefficients for each $x$, $y$, and $z$ component are 0.44, 0.98, and 0.86, respectively. The strong positive correlation confirmed the sunward nature of the AWs in the studied ICME's MC region.


 \subsection{Characteristics of the Alfv\'en Waves}
Generally, Els\"asser variables are used to characterize the solar-wind turbulence and AW properties \citep{elsasser1950hydromagnetic,marsch1987ideal,bruno2013solar}. Here, we employed them to distinguish the dominant flow of outward and inward Alfv\'enic fluctuations \citep{elsasser1950hydromagnetic, marsch1987ideal,bruno2013solar}. The Els\"asser variables are defined as %
\begin{equation} \label{zpm}
	\vec{Z}^{\pm} = \vec{V} \pm \frac{\vec{B}}{\sqrt{4\pi \rho}} 
\end{equation}
Here, $\vec{V}$  and $\vec{B}$ are fluctuations in the proton velocity and magnetic field respectively. The $\pm$ sign in front of $\vec{B}$ depends on the sign of [$-k ~\cdotp B_0$]. where \textit{\textbf{k}} is a wave vector. If both, the velocity and magnetic field are directed outward, Equation \ref{zpm} manifests as $\vec{Z^{+}}=\vec V -\vec V_A$ and $\vec{Z^{-}}=\vec V +\vec V_A$. On the other hand, if the magnetic field points in the direction towards the Sun, the correlation sign is reversed (\textit{\textbf{V}} points always outwards), and $\vec{Z^{+}}=\vec V +\vec V_A$ and $\vec{Z^{-}}=\vec V -\vec V_A$ are the results. In this sense, $Z^+$ and $Z^-$  represent  outward and inward  Alfv\'enic mode respectively, at all times. \citep{roberts1987origin, bruno1991origin, d2015origin}

\begin{figure}
	\centering
	\includegraphics[width = \columnwidth]{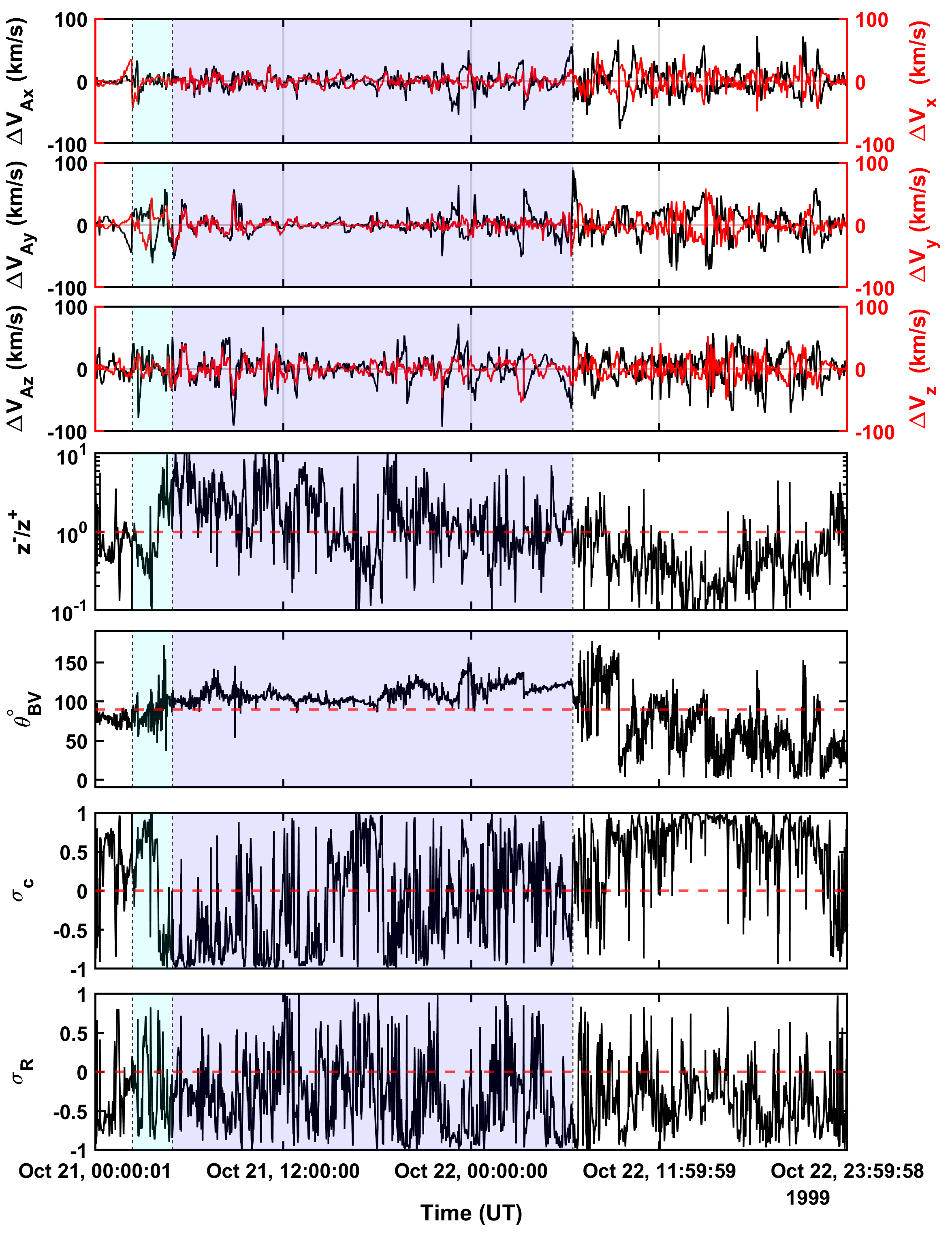}
	\caption{The top three panels compare Alfv\'en velocity fluctuations $\Delta V_{Ai}$ (red) to proton flow velocity fluctuations $\Delta V_p$ over time (blue). They demonstrate the Alfv\'enic features in MC and HSS regions. The ratio of Els\"asser variables ${z^-}/{z^+}$ is shown in the fourth panel. The fifth panel depicts the angle between the Alfv\'en velocity and the solar-wind speed. The bottom two panels show the temporal variation of the normalized cross helicity ($\sigma_c$) and normalized residual energy ($\sigma_R$ ). WIND (MFI and 3DP) spacecraft data are used for the above analysis with a time cadence of 3 sec.}
	\label{fig:Alf_label}
\end{figure}

Figure~\ref{fig:Alf_label} represents the characteristics of the AWs in question. The top three panels clearly show that components of $ \Delta V$ and $ \Delta V_A$ are either correlated or anti-correlated. This implies that the existence of AWs in the studied region exhibits both outward and inward propagation nature. In order to  gain a better clarity in the time evolution of outward and inward propagation of the waves, we have demonstrated the ratio  $Z^-/Z^+$, normalized cross helicity ($\sigma_c$), the angle between $ \Delta {V}$ and $\Delta {B}$ ($\theta_{VB} $), and the normalized residual energy ($\sigma_R$) in Figure~\ref{fig:Alf_label}. 

We find a mean value of the ratio of Els\"asser variables of $\sim 0.54$ in the HSS region whereas it is $\sim 2.13$ inside the ICME's MC. In fact, the temporal fluctuation in the ratio reached $\sim 10$ in the anterior part of the MC and highly varies in the trailing part of the MC.


The normalized cross helicity ($\sigma_c$) is defined as,
\begin{equation} 
	\sigma_c = (\frac{{e^+}-{e^-}}{{e^+}+{e^-}})
\end{equation} 
where $e^{\pm}$=$\frac{1}{2}$ $(z^{{\pm}})^2$, $e^-$ and $e^+$ are the energies related to $z^-$ and $z^+$. The normalized cross helicity shows the degree of Alfv\'enicity  \citep{tu1989basic}. Moreover, $\sigma_c \sim 1$  denotes a prominent outward propagating flow whereas $\sigma_c \sim -1$ depicts a dominating inward propagating flow \citep{matthaeus1982stationarity,tu1989basic}. The analysis yielded a positive average value for the HSS region, i.e. $\sigma_c~= ~0.57$, indicating predominately outward flow. Furthermore, we observed  $\sigma_c =-1$ with high fluctuations in the front part of the MC. Moreover, the mean value for the ICME's MC is found as $\sigma_c~ = ~-0.22$, implying inward flow in the MC region.

To quantify the wave propagation direction, we have estimated the angle ($\theta_{VB}$) between $V$ and $B_{\rm mag}$ as follows:
\begin{equation}
	\theta_{VB}=\cos^{-1}(\frac{-B_x}  {B_{\rm mag}})
\end{equation}
We frequently observed values of $\theta$ below 30$^\circ$, and the mean value around 40.21$^{\circ}$ in the HSS region. This implies the two vectors are nearly parallel in the HSS region, supporting the observed strong outward Alfv\'enic flow. However, in the MC region, we observed highly fluctuating angles, which sometimes reached a value of 150$^\circ$ . The mean value of the angle is observed at 109.67$^{\circ}$.

The normalized residual energy is defined as 
\begin{equation}
	\sigma_R=(\frac{e^v-e^b}{e^v+e^b})
\end{equation} 
where $e^v$ and $e^b$ are the kinetic and magnetic energies respectively. $\sigma_R$ is a measure of the excess magnetic field energy with respect to the kinetic energy or vice versa \citep{bruno2013solar}. Our analysis revealed that the value of $\sigma_R$ is routinely  below 1 in the HSS region, whereas it is highly fluctuating in the MC region, possibly due to the mixing of inward and outward waves.


 \section*{Discussion and Conclusion}
 
Alfv\'enic fluctuations are transverse magnetohydrodynamic (MHD) fluctuations in which ions and magnetic fields oscillate at low frequencies \citep{Cross1988, Cramer2001}. They propagate in the direction of the magnetic field, with the ion mass density providing inertia and the magnetic field lines providing a restoring tension force. Alfv\'enic fluctuations are ubiquitous in space plasmas, such as: the ionosphere, the magnetotail \citep{keiling2009alfven}, the magnetosheath, the interplanetary space \citep{wang2012large}, the slow solar wind \citep{d2019slow} and fast solar wind \citep{hollweg1975alfven,tsurutani2018review}, co-rotating interaction region (CIR) \citep{tsubouchi2009alfven,shi2020propagation},  the ICME's sheath \citep{shaikh2019coexistence} and MC \citep{raghav2018torsional},  the planetary region \citep{hinton2019alfven}, the inner-heliosphere \citep{bavassano1989large,perrone2020highly}, the outer-heliosphere, the  astrophysical plasma, the solar corona \citep{tomczyk2007alfven,cally2017alfven}, the solar surface or atmosphere \citep{jess2009alfven,mathioudakis2013alfven}, the lab-plasma \citep{Gekelman1999, Gekelman2003}. \cite{raghav2022first} found the existence of surface AWs in the ICME flux rope. The AW shows some peculiar characteristics such as period doubling phenomena, arc polarization and phase steepening \citep{Riley1996,tsurutani2018review, shaikh2019coexistence}. 

It is worth noting that the outward AWs are widespread in the solar wind, whereas inward AWs are uncommon \citep{belcher1969large,daily1973alfven,burlaga1976microscale,riley1996properties,yang2016observational}. With growing heliocentric distance, inward AWs are expected. They're also associated with unusual occurrences like magnetic reconnection exhausts and/or back-streaming ions from reverse shocks \citep{belcher1971large,roberts1987origin,bavassano1989large,gosling2009one,gosling2011pulsed}. Localized superposition of inward and outward AWs may be caused by the solar-wind velocity shear effect, triggered by plasma instabilities \citep{bavassano1989evidence}. When both AWs are present simultaneously, non-linear interactions occur \citep{dobrowolny1980fully}, which are essential for the dynamical evolution of a Kolmogorov-like MHD spectrum \citep{bruno2013solar}. In general, the solar wind is uniform and persistent in high latitudes; therefore a decrease in the cross helicity could be induced by parametric instability \citep{malara2001nonlinear}. The helicity decreases as the heliocentric distance increases \citep{bavassano2000alfvenic, matthaeus2004transport}.

Here, we demonstrated the existence of an AW during the ICME-HSS interaction at 1 AU. The observations in Figure \ref{fig:IP_label} indicate that the HSS interacts with the ICME from the trailing edge. In the studied ICME event, we observed the upstream shock with small magnetic field amplitude and very weak sheath region formation. The proton speed of this ICME has diminished in the leading part of the MC region. As a result, the ICME MC has similar characteristics to a slow MC. In the typical case, as referred to above, there would be no forward shock and sheath \citep{tsurutani2004properties}. In our case, there is a forward shock and weak/no sheath. But, we observed a reverse shock at the end of the ICME interval \citep{tsurutani2011review}. The ICME's velocity increased from $\sim 460$  km s$^{-1}$ to $\sim 700$  km s$^{-1}$, indicating that the HSS is dynamically compressing the ICME. The compression intensifies the magnetic field to almost double in magnitude in its rear part, resulting in a strong geomagnetic perturbation \citep{dal200617,singh2009geoeffectiveness,kilpua2012relationship}. As a result, the ICME's MC does not appear to be expanding as expected; rather, the rise in the total IMF  near the rear end implies the compression of the MC. In the $B_y$, $B_z$, and $\theta$ variations, the distortion is clearly visible \citep{kilpua2012relationship}. Based on the observations and estimations  shown in Figure \ref{fig:wal_label} and \ref{fig:Alf_label}, we explicitly observed the AW during this interaction. From the correlation values and temporal fluctuations in various estimated quantities using Els\"asser variables, we infer that the initial part of the ICME's MC superposed with a strong inward AW flow, whereas the HSS region displayed a strong outward flow. 


 
 \begin{figure}
 	\centering
 	\includegraphics[width = \columnwidth]{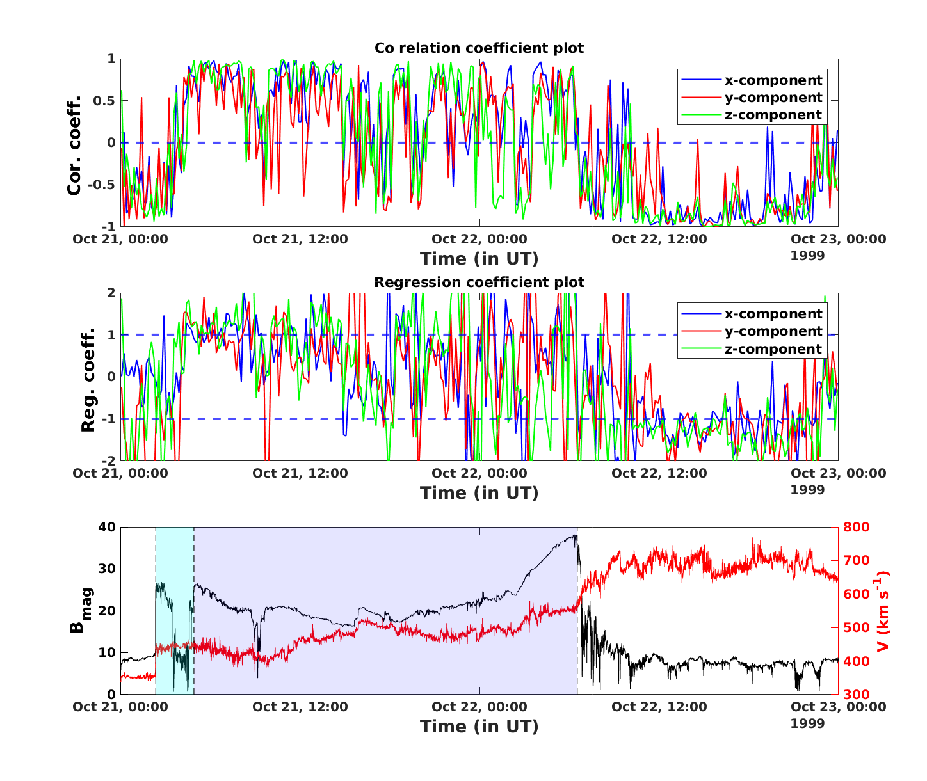}
 	\caption{The top and middle panel shows the fluctuations in correlation coefficient and the regression coefficient of each component of velocity and magnetic field . The bottom panel references the evolution of the ICME and HSS using $B_{mag}$ and solar wind speed. }
 	\label{fig:Alf_Rw}
 \end{figure}

 Alfv\'enic fluctuations in the solar wind are usually a mix of two populations: outward-propagating and inward-propagating \citep{d2015origin}. The Wal\'en slope (or the correlation between the magnetic field and plasma velocity) of AWs observed in the solar wind can be significantly reduced by a mix of inward and outward AWs \citep{belcher1971large,marsch1993modeling,bruno2013url, yang2016observational}.
 The observed AWs lie in the frequency range $10^{-3}$ to $10^{-1}~$Hz. Therefore, to verify the temporal variation of correlation coefficient and regression coefficient,  we passed the data of $V$ and $V_A$ components through the $4^{th}$ order Butter-worth MATLAB filter algorithm with a single broadband frequency boundaries of $10^{-3}$ to $10^{-1}~$Hz. We divided the data set being analyzed into ten-minute intervals (200 data points in each interval). Furthermore, we estimated the correlation coefficient and regression coefficient between the respective components of $V$ and $V_A$ for each time window. Figure~\ref{fig:Alf_Rw} shows the temporal variation of the correlation coefficient (top panel) and regression coefficient (middle panel) for the observed Alfv\'enic region. The correlation coefficient and regression coefficient fluctuate to $\sim -1$ in the HSS region, corroborating strong outward flow. In contrast, both quantities fluctuate to $\sim 1$ in the front part of the MC, suggesting the inward flow of Alfv\'enic fluctuations. However, we found highly fluctuating values for both coefficients and Els\"asser variables (see Figure \ref{fig:Alf_label}) at the trailing part of the MC. In the studied interaction case, the MC's internal (magnetic) pressure increases due to the compression exerted by the HSS. The resulting force sweeps the plasma backward, i.e., the reflection of ions from the rear boundary. Note that the AW's amplitude is lower in the MC's trailing part than in the HSS and front part. Therefore, we believe that the outward and inward AWs are generated or reflected at the rear boundary of the MC. Moreover, the mixing of inward and outward AWs within the trailing part of the MC region is possible, as suggested in the reported studies \citep{d2015origin}. 
 It is exciting to examine the inward-outward interaction region to understand parametric instabilities, which will be studied in the near future.


The generation of the AWs in our study could be attributed to the following reasons: (1) steepening of a magnetosonic wave that generates the shock at the leading edge of the MC \citep{tsurutani1988origin,tsurutani2011review} (2) The velocity shear due to the interaction between the ICME and HSS \citep{bavassano1978local,roberts1992velocity,hollweg2011alfven}. (3) \cite{tsurutani1988origin} distinguished the driver gas/MC as regions without AWs or discontinuities. But, AWs are detected, implying that they may be a leak in the MC from the HSS or generated locally. (4) Besides, there is a reverse shock observed on $24^{th}$ October 1999, which may have some role in the AWs being generated through the backstreaming of ions \citep{gosling2011pulsed}.
(5) Furthermore, the reconnection region observed at the leading edge of the MC \citep{petschek196450,levy1964aerodynamic,gosling2005magnetic}, and \cite{gosling2005magnetic} suggest that the reconnection exhaust at the heliospheric current sheet (HCS) can generate AWs. 
(6) Apart from this, a simulation study by \cite{tsubouchi2009alfven}  claimed that the Alfv\'enic fluctuations in a HSS interact with a velocity gradient structure. The initial AWs break into two Alfv\'en modes that travel in opposite directions. Here, powerful parallel and antiparallel flows produce the field gradient at the edges, acting as a mirror force to modify the magnetic intensity.	
Moreover, \cite{wang2019multispacecraft} perform multi-spacecraft observations of the MC and suggest that AWs can have  unidirectional as well as bidirectional AWs. They also speculated that unidirectional AWs are formed within an MC by distortions in a pre-existing flux rope, while bidirectional AWs are emitted from the centre of reconnection and subsequently move outward along two-foot legs of an ICME flux rope. In our study, it appears that the AWs (inward waves) within the MC could be due to distortions of the MC. This distortion is caused by the HSS. Furthermore, this AW embedded MC travels in the sea of the HSS, which has an outward AW. Thus, our study will be significant in understanding the ICME-solar wind interaction and underlying physical mechanism of these waves.



\section*{Acknowledgements}
The authors would like to acknowledge all individuals involved with WIND spacecraft mission development, data providing team, etc. The authors thank Mr. Greg Hilbert for valuable suggestion.  We also acknowledge the NASA/GSFC Space Physics Data Facilities (CDAWeb or ftp) service. 


\bibliographystyle{unsrtnat}
\bibliography{main}  






\end{document}